\documentclass[12pt,letterpaper]{article}
\usepackage[latin1]{inputenc}
\usepackage{amsmath}
\usepackage{amsfonts}
\usepackage{amssymb}
\usepackage{graphicx}
\usepackage{caption}
\usepackage{endnotes}

\begin{document}
\title{Portfolio Optimization Under Uncertainty}
\author{Alex Dannenberg\footnote{alex.dannenberg@pinemountaincapital.com}\\
		Pine Mountain Capital Management\\
		August 2009
		}
\date{}		
\maketitle
\section{Abstract}
Classical mean-variance portfolio theory\endnote{H. Markowitz, \textit{Portfolio Selection: Efficient Diversification of Investments} (1959)}\endnote{W. Sharpe, \textit{Portfolio Theory and Capital Markets} (1970)} tells us how to construct a portfolio of assets which has the greatest expected return for a given level of return volatility.  Utility theory then allows an investor to choose the point along this efficient frontier which optimally balances her desire for excess expected return against her reluctance to bear risk.  The means and covariances of the distributions of future asset returns are assumed to be known, so the only source of uncertainty is the stochastic piece of the price evolution. 
\vfill\vfill\noindent
In the real world, we have another source of uncertainty - we estimate but don't know with certainty the means and covariances of future asset returns.  This note explains how to construct mean-variance optimal portfolios of assets whose future returns have uncertain means and covariances.  The result is simple in form, intuitive, and can easily be incorporated in an optimizer.
\vfill\vfill\noindent
Various approaches already exist to improve portfolio construction in the presence of uncertain dynamics. Factor models\endnote{R. Grinold and R. Kahn, \textit{Active Portfolio Management} (1995)} and random matrix theory\endnote{Laloux, Cizeau, Bouchaud and Potters, Phys.Rev.Lett.83, 1467 (1999)}\endnote{Plerou, Gopikrishnan, Rosenow, Amaral and Stanely, Phys.Rev.Lett.83, 1471 (1999)} can be used to provide de-noised covariance and correlation matrices as inputs to optimizers, thereby ameliorating the effects of over-fitting.  They do not, however, allow one to correct for the effects of uncertainty in expected returns.  Re-sampled efficient frontiers\endnote{R. Michaud, \textit{Efficient Asset Management} (1998)} provide a reasonable, simulation-based way to assess the stability of a given portfolio's performance to sampling uncertainty - and re-sampled efficiency is a reasonable, if ad hoc, metric to consider when constructing a set of portfolio weights.  Other approaches exist as well , and I won't attempt to enumerate or summarize the long list.  But a tractable, closed-form, theoretically-grounded approach to incorporating the joint effects of uncertainty in expected returns and covariances is, to my knowledge, lacking.  This is odd because, as we shall see, it's not that hard\ldots
\section{A Brief Review}
We start by setting some notation and re-deriving the classical results:  An investor will invest a fraction $f_0$ of her wealth $W$ in a riskless asset and fractions $f_{i\in\lbrace1:N\rbrace}$ in each of $N$ risky assets.  The risky assets' prices $P_i$ are assumed to undergo known covarying diffusions
\[\frac{{d{P_i}}}
{{{P_i}}} = {\mu _i}({P_i})\,dt + {\sigma _i}({P_i})\,d{z_i},\,\,\left\langle {d{z_i}d{z_j}} \right\rangle  = {\rho _{ij}}\;dt\]
so that
\[dW = ({f_0}W)\,r\,dt + \sum\limits_{i = 1}^N {({f_i}\,W)\,\frac{{d{P_i}}}
{{{P_i}}} = W\,[r\,dt + \sum\limits_{i = 1}^N {f{\,_i}({\mu _i} - r)\,dt + {f_i}\,{\sigma _i}\,d{z_i}]} } \]
and the investor is assumed to choose her investment fractions so as to maximize the expected change in her utility\footnote{We use a power-law utility function for convenience only.  Other choices of utility function with positive first derivative and negative second derivative simply lead to a different risk-aversion constant in the expression for $Q$ below.  The analysis goes through without change.}, $U$, where 
\[{U_t} = U({W_t}) \equiv \frac{1}
{x}({W_t}^x - 1),\,\,x < 1.\]
Noting that $U^{\prime} = {W^{x - 1}} > 0$ and that $U^{\prime\prime} = (x - 1)\,{W^{x - 2}} < 0$, Ito's lemma gives us
\[\left\langle {d{U_t}} \right\rangle  = {W_t}^x\,[\,r + \sum\limits_{i = 1}^N {{f_i}\,({\mu _i} - r) + \frac{{x - 1}}
{2}\sum\limits_{i,j = 1}^N {{f_i}\,{f_j}\,{\sigma _i}\,{\sigma _j}\,} } {\rho _{ij}}]\,dt.\]
Thus we want to choose the $f_i$ to maximize a quantity
\[Q \equiv \sum\limits_{i = 1}^N {{f_i}\,({\mu _i} - r) + \frac{{x - 1}}
{2}\sum\limits_{i,j = 1}^N {{f_i}\,{f_j}\,{\sigma _i}\,{\sigma _j}\,} } {\rho _{ij}}\]
Note that we expect an investment fraction, a dimensionless quantity, to be on the order of 
$\frac{{\mu  - r}}
{{{\sigma ^2}}}$
since this is the simplest dimensionless quantity we can form from the parameters that describe the dynamics (only excess return is relevant to investment in a risky asset).  Indeed, in a world with only one risky security it is the case that the optimal investment fraction is exactly 
$\frac{{\mu  - r}}
{{{\sigma ^2}}}$.  
We will see that the optimization problem is greatly simplified by denominating our investment fraction in these units\ldots  We define
\[{c_i}:\;\;{f_i} = {c_i}\frac{{{\mu _i} - r}}
{{\sigma _i^2}}\]
so that
\[\begin{gathered}
  Q = \sum\limits_{i = 1}^N {{c_i}\,{{(\frac{{{\mu _i} - r}}
{{{\sigma _i}}})}^2}}  + \,\frac{{x - 1}}
{2}\sum\limits_{i,j = 1}^N {{c_i}\,{c_j}\,} (\frac{{{\mu _i} - r}}
{{{\sigma _i}}})\,(\frac{{{\mu _j} - r}}
{{{\sigma _j}}})\,{\rho _{ij}} \hfill \\
  \;\;\; = \sum\limits_{i = 1}^N {{c_i}\,{S_i}^2}  + \,\frac{{x - 1}}
{2}\sum\limits_{i,j = 1}^N {{c_i}\,{c_j}\,} {S_i}\,{S_j}{\rho _{ij}} \hfill \\ 
\end{gathered} \]
where $S$ denotes Sharpe Ratio.  Defining the $N\times{N}$ matrix
\[\Phi :{\Phi _{ij}} \equiv {S_i}\,{S_j}\,{\rho _{ij}}\]
and the $N\times1$ vector
\[\vec \Delta :{\Delta _i} \equiv {S_i}^2 = {\Phi _{ii}}\]
we can write
\[Q = \vec{c}\,^t\,\vec \Delta  + \frac{{x - 1}}
{2}\vec{c}\,^t\,\Phi\,\vec{c}.\]
If all drifts and covariances are known with certainty, we can maximize $Q$ over the $c_i$ as per usual:
\[0 = ({\partial _{{c_i}}}Q\left| \;{\vec c = {{\vec c}\,^*})} \right. = {\Delta _i} + (x - 1)\,{(\Phi {\vec c\,^*})_i},\forall \;i \in [1,N]\]
which implies that
\[{\vec c\,^*} = \frac{1}
{{1 - x}}{\Phi ^{ - 1}}\vec \Delta \]
or, equivalently,
\[{f_i}^* = \frac{1}{{1 - x}}[{\Phi ^{ - 1}}\vec \Delta ){]_i}\frac{{{S_i}}}{{{\sigma _i}}}\]
which is the standard Markowitz result in our notation.
\\\\
\noindent
It's worth noting what our choice of variables has gained us:  The previous equation makes transparent the fact that, in order to maximize her (change in) utility, our classical mean-variance optimal investor should allocate her risk budget among risky assets ``in proportion to'' their Sharpe ratios.  Specifically, if she's choosing how to spread her capital among $N$ uncorrelated assets then ${\Phi ^{ - 1}}\vec \Delta  = {\rm I}$ and she'll invest a fraction ${f_i}^*$ of her capital in the i-th risky asset so that the risk, ${f_i}^*{\sigma _i}$, is equal to $\frac{{{S_i}}}
{{1 - x}}$.  Nothing new so far\ldots
\section{Introducing Parameter Uncertainty}
But what if the drifts and covariances are themselves uncertain?  Now our investor's future utility is uncertain not only because of the noise intrinsic to the risky assets, but also because her ability to characterize those risky assets is imperfect.  This is a bad thing for her because ${\partial ^2}\left\langle {d{U_t}} \right\rangle /\partial {f_i}\partial {f_j}$ is negative (recall $x<1$), and therefore $\left\langle {d{U_t}} \right\rangle $ declines as the uncertainty of her estimated $f_i$ increases - even if she has the right values on average.\footnote{Just as the positive convexity of an option's payout with respect to the price of the underlying asset gives rise to time-value if future prices are uncertain\ldots  Note that this same negative convexity justifies fractional Kelley strategies in the realm of proportional betting systems.}  We'll choose portfolio weights so as to maximize her expected future utility, where the expectation is taken over \textit{both} the return uncertainty for a given process \textit{and} the parameter uncertainty for that process.  Let's define ${\hat \mu _i}$ to be her estimate of ${\mu _i}$, the true expected return for asset $i$.  Similarly, let ${\hat \sigma_i},\,{\hat \rho_{ij}}$ and ${\hat S_i}$ be the estimated values of ${\sigma_i},\,{\rho_{ij}}$ and ${S_i}$.  Finally, we'll define 
\[{\hat c_i}:{f_i} = {\hat c_i}\frac{{{{\hat \mu }_i} - r}}{{\hat \sigma _i^2}}\]
so that 
\[{f_i} = {c_i}\frac{{{\mu _i} - r}}{{\sigma _i^2}} = {\hat c_i}\frac{{{{\hat \mu }_i} - r}}{{\hat \sigma _i^2}}.\]
Note that the $c_i$ and $\hat{c_i}$ simply represent the same real-world investment fractions $f_i$ in different units.  This means that
\begin{eqnarray*}
Q &\equiv& \sum\limits_{i = 1}^N {{c_i}\,{{(\frac{{{\mu _i} - r}}{{{\sigma _i}}})}^2}}  + \frac{{x - 1}}{2}\sum\limits_{i,j = 1}^N {{c_i}\,{c_j}} \,(\frac{{{\mu _i} - r}}{{{\sigma _i}}})\,(\frac{{{\mu _j} - r}}{{{\sigma _j}}})\,{\rho _{ij}}\\
&=& \sum\limits_{i = 1}^N {{{\hat c}_i}\,{{(\frac{{{{\hat \mu }_i} - r}}{{{{\hat \sigma }_i}}})}^2}\,\frac{{{\mu _i} - r}}{{{{\hat \mu }_i} - r}}} \,\, + \,\,\frac{{x - 1}}{2}\sum\limits_{i,j = 1}^N {{{\hat c}_i}\,{{\hat c}_j}\frac{{{{\hat \mu }_i} - r}}{{{{\hat \sigma }_i}}}} \,\,\frac{{{{\hat \mu }_j} - r}}{{{{\hat \sigma }_j}}}{\hat \rho _{ij}}\,\,\frac{{{\sigma _i}}}{{{{\hat \sigma }_i}}}\,\,\frac{{{\sigma _j}}}{{{{\hat \sigma }_j}}}\,\,\frac{{{\rho _{ij}}}}{{{{\hat \rho }_{ij}}}}\\
&=& \sum\limits_{i = 1}^N {{{\hat c}_i}\,\;{{\hat S}_i}^2\;\,\frac{{{\mu _i} - r}}{{{{\hat \mu }_i} - r}}} \,\, + \,\,\frac{{x - 1}}{2}\sum\limits_{i,j = 1}^N {{{\hat c}_i}\,\;{{\hat c}_j}\;{{\hat S}_i}\;{{\hat S}_j}\;} {\hat \rho _{ij}}\,\,\frac{{{\sigma _i}}}{{{{\hat \sigma }_i}}}\,\,\frac{{{\sigma _j}}}{{{{\hat \sigma }_j}}}\,\,\frac{{{\rho _{ij}}}}{{{{\hat \rho }_{ij}}}}
\end{eqnarray*}
It's important to note that any term in $Q$ involving $\hat\mu_i$ is zero when $\hat\mu_i=r$, which implies that
\[\int\limits_{ - \infty }^{ + \infty } {d{\hat\mu}\;{\rm P}{{(\hat \mu )}^{}}\;Q(\hat \mu )}  = \mathop {\lim }\limits_{\varepsilon  \to 0} (\int\limits_{ - \infty }^{r - \varepsilon } { + \int\limits_{r + \varepsilon }^\infty  {} {)^{}}{d{\hat\mu}}\;{\rm P}{{(\hat \mu )}^{}}\;Q(\hat \mu )} \]
This allows us to write
\begin{eqnarray*}
\left\langle Q \right\rangle  &=& {\sum\limits_{i = 1}^N {{{\hat c}_i}\;\,{{\hat S}_i}^2\;\,\left\langle {\frac{{{\mu _i} - r}}{{{{\hat \mu }_i} - r}}} \right\rangle } _{{{\hat \mu }_i} \ne r}}\; + \;\frac{{x - 1}}{2}\sum\limits_{i,j = 1}^N {{{\hat c}_i}\,\;{{\hat c}_j}\;{{\hat S}_i}\;{{\hat S}_j}} \,\,{\hat \rho _{ij}}\,\left\langle {\frac{{{\sigma _i}}}{{{{\hat \sigma }_i}}}\,\,\frac{{{\sigma _j}}}{{{{\hat \sigma }_j}}}\,\,\frac{{{\rho _{ij}}}}{{{{\hat \rho }_{ij}}}}} \right\rangle\\
&=& {\vec{\hat{c}}\,^t}(\vec{\hat{\Delta}}  * \vec A) + \frac{{x - 1}}{2}\,\,{\vec{\hat{c}}\,^t}(\hat \Phi  * B)\,\,\vec{\hat{c}}\\
\rm{where}\\\\
\hat \Phi :{\hat \Phi _{ij}} &=& {\hat S_i}\;{\hat S_j}\;{\hat \rho _{ij}}\\
\vec{\hat{\Delta}} :{\hat \Delta _i} &=& {\hat S_i}^2 = {\hat \Phi _{ii}}\\
\vec A:{A_i} &=& {\left\langle {\frac{{{\mu _i} - r}}{{{{\hat \mu }_i} - r}}} \right\rangle _{{{\hat \mu }_i} \ne r}}\\
B:{B_{ij}} &=& \left\langle {\frac{{{\sigma _i}}}{{{{\hat \sigma }_i}}}\,\,\frac{{{\sigma _j}}}{{{{\hat \sigma }_j}}}\,\,\frac{{{\rho _{ij}}}}{{{{\hat \rho }_{ij}}}}} \right\rangle
\end{eqnarray*}
and $*$ denotes element-by-element multiplication (Matlab's .* operator or R/S-plus's * operator).  Note that we don't need to address the presence of the pole in $\left\langle\frac{\mu_i - r}{\hat{\mu}_i - r}\right\rangle$ because the evaluation of  $\left\langle Q \right\rangle$ requires only  $\,{\left\langle {\frac{\mu_i - r}{\hat{\mu}_i} - r} \right\rangle _{\hat{\mu}_i \ne r}}$.  Maximizing $\left\langle Q \right\rangle$ over the $\hat{c}_i$ gives
\[\vec{\hat{c}}^* = \frac{1}{1 - x}{(\hat{\Phi}  * B)^{ - 1}}(\vec{\hat{\Delta}}  * \vec{A})\]
or, equivalently,
\[{f_i}^* = \frac{1}{{1 - x}}{[{(\hat \Phi  * B)^{ - 1}}(\vec{\hat{\Delta}}  * \vec A)]_i}\frac{{{{\hat S}_i}}}{{{{\hat \sigma }_i}}}.\]
Plugging in these optimal investment fractions gives
\[\left\langle Q \right\rangle  = \frac{1}{2}\;\frac{1}{{1 - x}}\;{(\vec{\hat{\Delta}}  * \vec A)^t}\;(\hat \Phi  * B){\,^{ - 1}}\;(\vec{\hat{\Delta}}  * \vec A)\]
or
\[{\left\langle {d{U_t}} \right\rangle ^*} = dt\;\;{W_t}^x\,[\,r + \frac{1}{2}\;\frac{1}{{1 - x}}\;{(\vec{\hat{\Delta}}  * \vec A)^t}\;(\hat \Phi  * B){\,^{ - 1}}\;(\vec{\hat{\Delta}}  * \vec A)]\,.\]
\textbf{This formula for $\mathbf{{f_i}^*}$ is our main result.}  Note that when there's no uncertainty in expected returns or covariances then ${A_i} = {B_{ij}} = 1,\forall i,j$ and we regain the standard answer.  A bit more generally, if ${A_i} = a,\forall i$ and ${B_{ij}} = b,\forall i,j$, i.e. if relative uncertainties in drift and volatility are assumed to be the same across all securities, then ${\vec{\hat{c}}^*}$ is equal to $\frac{a}{b}$  times the standard result - which just corresponds to a change of leverage.  Since most real-world investors specify their risk-tolerance exogenously and use optimizers only to determine relative position sizes, the above prescription for incorporating parameter uncertainty has no real effect in this circumstance.  In other words, the prescription is likely to be useful only when relative uncertainties in expected return and volatility differ across securities and/or uncertainties in correlations are introduced.

\section{Practical Application}
This is all fine and dandy, you say, but what should I actually \textit{do}?  How can one quantify these $\vec{A}$ and $B$ objects?  Well, if an investor's estimation procedure is quantitative then the fitting procedure might return covariances for the estimated parameters that allow $\vec{A}$ and $B$ to be computed directly.  If not, then simple tools can allow an investor to quantify her degree of certainty.  In the following paragraphs I describe a few reasonable but ad hoc parameterizations that allow closed form expressions for $\vec{A}$ and $B$ (ignoring correlations among uncertainties).  Plots of these functions with ``sliders'' for the input parameters can be built as part of an optimizer's GUI to enable the non-quantitative portfolio manager to include the uncertainty of her estimates, as well as the estimates themselves, in the portfolio construction process.
\\\\
\noindent
The remainder of this paper is organized as follows:  First we propose simple and reasonable distributions that an investor can use to describe her uncertainty about expected returns, volatilities and correlations.  Then we explain how to compute $\vec{A}$ and $B$ using these distributions.  Having set the stage, we conclude with a detailed, quantitative example.
\\\\
\noindent
\mathversion{bold}
\textbf{Estimate of} $ < \frac{{\mu  - r}}{{\hat \mu  - r}} > {_{\hat \mu  \ne r}}:$
\mathversion{normal}
\\\\Example 1:  Let's assume we have an unbiased estimate of the expected return of the i-th security $\hat{\mu_i} = \mu_i + x_i,\,\,x_i\thicksim N(0,\Sigma_i)$ so that $\left\langle {{{\hat \mu }_i}} \right\rangle  = {\mu _i}$.  This allows us to write
\[{A_i} = {\left\langle {\frac{1}{{1 + y_i}}} \right\rangle _{{y_i} \ne  - 1}}\,\,\,y_i\thicksim N(0,\frac{\Sigma _i}{\mu _i - r})\]
Note that $\frac{{{\Sigma _i}}}{{{\mu _i} - r}}$ is just the relative uncertainty in the estimate of the expected excess return of the i-th security.  A simple exercise in Gaussian integral evaluation gives us:
\[\begin{array}{l}
 A = {\left\langle {\frac{1}{{1 + y}}} \right\rangle _{y \ne  - 1}} = \mathop {\lim }\limits_{\varepsilon  \to 0} (\int\limits_{ - \infty }^{ - 1 - \varepsilon } { + \int\limits_{ - 1 + \varepsilon }^\infty  {} )\frac{{dy}}{{\sqrt {2\pi {\sigma ^2}} }}} \frac{{{e^{ - {y^2}/2{\sigma ^2}}}}}{{1 + y}},\,\,\,\sigma  \equiv \frac{\Sigma }{{\mu  - r}} \\ 
 \,\,\,\,\,\,\,\,\,\,\,... = \frac{{{e^{ - 1/2{\sigma ^2}}}}}{{{\sigma ^2}}}\sum\limits_{n = 0}^\infty  {\frac{1}{{n!\,{2^n}\,(2n + 1)\,{\sigma ^{2n}}}}}  \\ 
 \end{array}\]
This series converges quite quickly, even for unrealistically small values of $\sigma$, i.e. even when there's a high degree of certainty that $\hat\mu$ is close to the true value of $\mu$.  The resulting function $A(\sigma)$ looks like this:
\newpage
\begin{figure}[!htbp]
\centering
\includegraphics[width=0.85\textwidth, height = 0.85\textwidth]{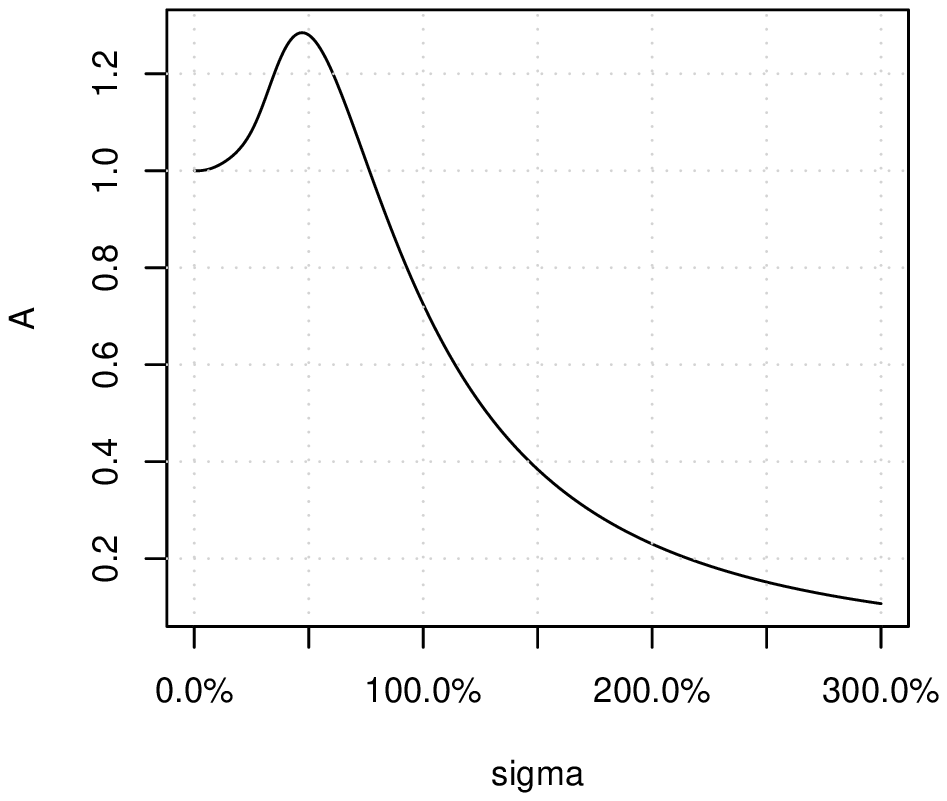} %
\end{figure}

This shape is easy to understand qualitatively:  If the uncertainty in $y_i$ is zero then $y_i$ is equal to its mean value of zero and $A_i=1$.  For values of $\frac{{{\Sigma _i}}}{{{\mu _i} - r}}$ small enough that there's negligible probability of ${y_i} \le  - 1$ then the positive convexity of $\frac{1}{{1 + {y_i}}},{y_i} >  - 1$ ensures that $\left\langle {\frac{1}{{1 + {y_i}}}} \right\rangle  > (\frac{1}{{1 + \left\langle {{y_i}} \right\rangle }} = 1)$.  But for very large values of $\frac{{{\Sigma _i}}}{{{\mu _i} - r}}$, $P(y_i)$ becomes so broad that $\left\langle {\frac{1}{{1 + {y_i}}}} \right\rangle  \approx \left\langle {\frac{1}{{{y_i}}}} \right\rangle  = 0$ by symmetry.
\\\\Example 2:  How should we compute $A_i$ if we're concerned that our forecast alphas are biased upward as a result of data mining?  In this case, it's instructive and qualitatively reasonable to assume that $\hat{\mu_i} = \mu_i + x_i,$ where $x_i\thicksim N(\frac{\mu_i - r}{2},\frac{3(\mu_i - r)}{4}).$  Note that we've ``deflated'' our expected returns but we're still ascribing predictive power to our data-mined results since an observation of ${\hat \mu _i} = r$ is deemed to be 2-sigma event.  Note, too, that this assumption fixes $A$:
\begin{eqnarray*}
A_i &=& \left\langle{\frac{\mu_i - r}{\hat{\mu_i} - r}}\right\rangle  = \left\langle{\frac{\mu_i - r}{\mu_i - r + x_i}}\right\rangle ,\;x_i\thicksim N(\frac{\mu_i - r}{2},\frac{3\;(\mu_i - r)}{4}) \\ 
    &=& \left\langle {\frac{1}{{1 + {y_i}}}} \right\rangle ,\;\;{y_i}\thicksim N(\frac{1}{2},\frac{3}{4}) \\ 
    &=& \frac{2}{3}\left\langle {\frac{1}{{1 + {z_i}}}} \right\rangle ,\;\;{z_i}\thicksim N(0,\frac{1}{2}) \\ 
    &=& \frac{2}{3}\;1.28 \\ 
    &=& 0.85. 
\end{eqnarray*}
\\\\
\noindent
\mathversion{bold}
\textbf{Estimate of} $\langle \frac{\sigma }{{\hat \sigma }}\rangle :$
\mathversion{normal}
A convenient way to parameterize our estimate of the i-th security's return volatility is ${\hat{\sigma}_i} = {\sigma _i}{e^{{x_i}}},{x_i}\thicksim N( - \frac{{{\Sigma _i}^2}}{2},{\Sigma _i})$ where I'm denoting the uncertainty by $\Sigma_i$ for reasons of convenience and familiarity, but the $\Sigma_i$ in this discussion of  is not related to the $\Sigma_i$ in the discussion of $\frac{{{\mu _i} - r}}{{{{\hat \mu }_i} - r}}$ above.  This parameterization ensures that $\left\langle {{{\hat \sigma }_i}} \right\rangle  = {\sigma _i},\,\,P({\hat \sigma _i} < 0) \equiv 0$ and allows us to compute $\left\langle {\frac{{{\sigma _i}}}{{{{\hat \sigma }_i}}}} \right\rangle  = {e^{{\Sigma _i}^2}} > 1$ and $\left\langle {\frac{{{\sigma _i}^2}}{{{{\hat \sigma }_i}^2}}} \right\rangle  = {e^{3{\Sigma _i}^2}}.$  $\Sigma_i$ can be chosen to represent the uncertainty of the volatility forecast using the relation ${\Sigma _i}^2 = \ln (1 + \frac{{{\mathop{\rm var}} ({{\hat \sigma }_i})}}{{{\sigma _i}^2}})$ or by fitting a stochastic volatility model.
\\\\
\noindent
\mathversion{bold}
\textbf{Estimate of} $\frac{\rho }{{\hat \rho }}:$
\mathversion{normal}
Dropping subscripts, we start as before by writing $\hat \rho  = \rho  + x,$ where    $x \in \left[ { - 1 - \rho ,\,1 - \rho } \right]$ has zero mean if $\hat\rho$ is assumed to be an unbiased estimator.  This gives $\left\langle {\frac{\rho }{{\hat \rho }}} \right\rangle  = \left\langle {\frac{1}{{1 + y}}} \right\rangle ,$ where $y \equiv \frac{x}{\rho } \in \left[ { - 1 - \frac{1}{{\left| \rho  \right|}}, - 1 + \frac{1}{{\left| \rho  \right|}}} \right]$ has zero mean.  Note that the fact that $y$ has zero mean but is distributed over a range symmetric about $-1$ implies that the pdf for $y$ will have positive skew.  We can learn quite a bit about the form of 
$\left\langle {\frac{\rho }{{\hat \rho }}} \right\rangle $ as a function of $\rho$ without doing any calculations:  To begin with, we know that symmetry requires that 
$\left\langle {\frac{\rho }{{\hat \rho }}} \right\rangle  \to 0$ as $\rho  \to 0$ (assuming $P(\hat \rho )$ unimodal).  We also expect that $\left\langle {\frac{\rho }{{\hat \rho }}} \right\rangle  \to 1$ as $\left| \rho  \right| \to 1$ because $P(\hat \rho ) \to \delta \left( {\hat \rho  \pm 1} \right)$ as $\left\langle {\hat \rho } \right\rangle  \to  \mp 1$.  Finally, we see that $\left\langle {\frac{\rho }{{\hat \rho }}} \right\rangle  > 1$ for $\left| \rho  \right| = 1 - \varepsilon $, $\left| \varepsilon  \right| <  < 1$ because $\left\langle {\frac{\rho }{{\hat \rho }}} \right\rangle  = \left\langle {\frac{1}{{1 + y}}} \right\rangle  = \left\langle {1 - y + {y^2} + h.o.} \right\rangle  \approx 1 + \left\langle {{y^2}} \right\rangle  > 1$ when the support for $y$ is near zero, as it is when $\left| {\hat \rho } \right| = 1 - \varepsilon $.  Thus, $\left\langle {\frac{\rho }{{\hat \rho }}} \right\rangle $ as a function of $\rho$ should behave qualitatively as follows:
\begin{figure}[!htbp]
\centering
\includegraphics[width=0.85\textwidth, height = 0.85\textwidth]{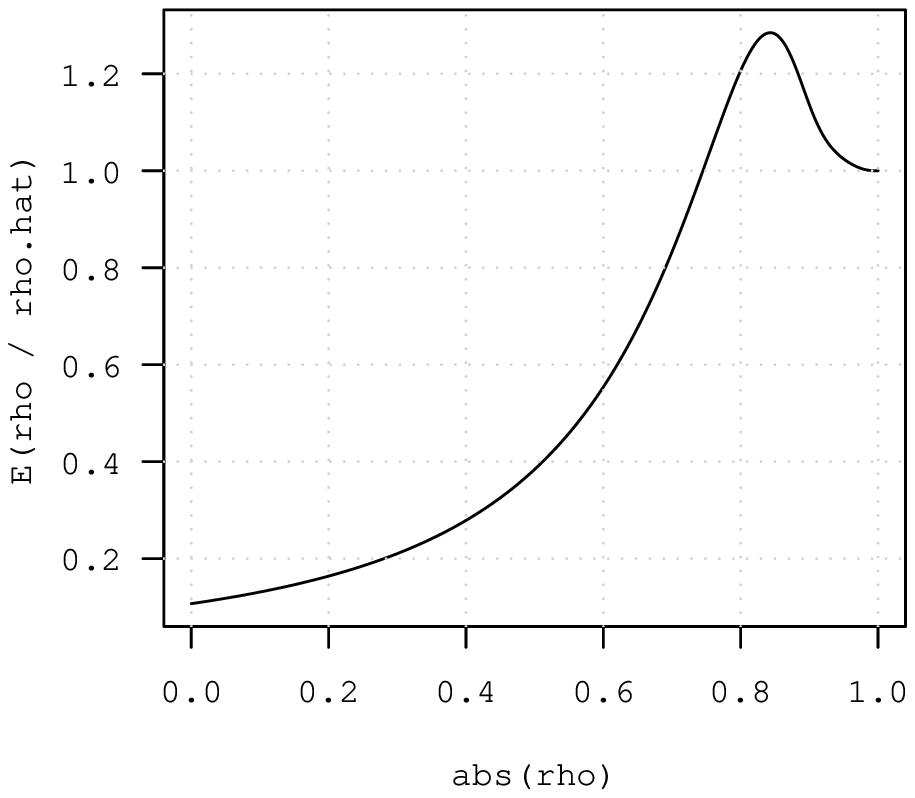} %
\end{figure}
\\A useful form for the probability distribution of $\hat\rho$ that's simple to work with is:
\[P(\hat \rho ) \propto {(1 - {e^{({{\hat \rho }^2} - 1)}})^\alpha }{(1 \pm \hat \rho )^{n - odd}}\]
The user needs to find two exponents that characterize her estimate, normalize the distribution, and use it to compute $\left\langle {\frac{\rho }{{\hat \rho }}} \right\rangle $.  By doing so, one can easily verify that the preceding figure is indeed a reasonable guide to the value of $\left\langle {\frac{\rho }{{\hat \rho }}} \right\rangle $ for a given value of $\left\langle\hat\rho\right\rangle $.
\section{A Detailed Example}
Consider a world with just two risky assets.  The (annualized) parameters that will govern the time evolution of their prices are:
\[\begin{array}{l}
 {\mu _1} = {\mu _2} = r + 10\% , \\ 
 {\sigma _1} = {\sigma _2} = 30\% , \\ 
 {\rho _{12}} = 0. \\ 
 \end{array}\]
These parameters are unobservable and can only be estimated.  If an investor \textit{did} know the true parameters, symmetry implies that she'd choose $f_1=f_2$ and achieve a (leverage-independent) Sharpe ratio = (expected excess return / return volatility) $=10\%$/$\frac{30\%}{\sqrt 2 }\approx 0.47$ at whatever leverage her utility function happened to dictate (the factor of $\sqrt 2$ is just the diversification effect of our assumption that the return processes are uncorrelated).  Any parameter uncertainty will reduce the Sharpe ratio to less than $0.47$, because ${\hat \mu _1} \ne {\hat \mu _2}{\rm{ and/or }}{\hat \sigma _1} \ne {\hat \sigma _2}$ will lead to a sub-optimal weighting in which ${f_1} \ne {f_2}$. 
\\\\\noindent
Of course, the parameters are \textit{not} known and must be estimated.  Let's assume that an investor's estimates will be unbiased and may be thought of as being drawn from the following sampling distributions:
\[\begin{array}{l}
 {{\hat \mu }_1} = {\mu _1} + {z_1},\,\,{z_1} \in N(0,\,\;5\% ),{\text{ so that stdev(}}\frac{{{{\hat \mu }_{\text{1}}}}}{{{\mu _{\text{1}}}}}{\text{)}} = {\text{ }}\frac{{5\% }}{{10\% }} = 0.5 \\ 
 {{\hat \mu }_2} = {\mu _2} + {z_2},\,\,{z_2} \in N(0,\;10\,\% ),{\text{ so that stdev(}}\frac{{{{\hat \mu }_{\text{2}}}}}{{{\mu _{\text{2}}}}}{\text{)}} = {\text{ }}\frac{{{\text{10\% }}}}{{{\text{10\% }}}} = 1 \\ 
 {{\hat \sigma }_1} = {\sigma _1}{e^{{x_1}}},\,\,{x_1} \in N( - 0.5\% ,\,10\% ){\text{, so that stdev(}}\frac{{{{\hat \sigma }_{\text{1}}}}}{{{\sigma _{\text{1}}}}}{\text{)}} \cong \frac{{3\% }}{{30\% }} = 0.1{\text{ and }}\left\langle {\frac{{{{\hat \sigma }_{\text{1}}}}}{{{\sigma _{\text{1}}}}}} \right\rangle  = 1 \\ 
 {{\hat \sigma }_2} = {\sigma _2}{e^{{x_2}}},\,\,{x_2} \in N( - 4.5\% ,\,30\% ){\text{, so that stdev(}}\frac{{{{\hat \sigma }_{\text{2}}}}}{{{\sigma _{\text{2}}}}}{\text{)}} \cong \frac{{9\% }}{{30\% }} = 0.3{\text{ and }}\left\langle {\frac{{{{\hat \sigma }_{\text{2}}}}}{{{\sigma _{\text{2}}}}}} \right\rangle  = 1 \\ 
 {{\hat \rho }_{{\text{12}}}} = {\rho _{{\text{12}}}} = 0\,\,{\text{(to prevent overcomplicating the example)}} \\ 
 \end{array}\] 
Recalling that 
\[{A_i} = \frac{{{e^{ - 1/(2{x_i}^2)}}}}{{{x_i}^2}}\sum\limits_{n = 0}^\infty  {\frac{1}{{n!\,{2^n}\,(2n + 1)\,{x_i}^{2n}}}} :\;{\rm{ }}{x_i} = \frac{{{\rm{stdev(}}{{\hat \mu }_{\rm{i}}})}}{{{{\hat \mu }_i}}},\]

\[{B_{ii}} = {e^{3{y_i}^2}}{\rm{and}}\;{B_{ij(i \ne j)}} = {e^{{y_i}^2 + {y_j}^2}}:\;\;\frac{{{\rm{var(}}{{\hat \sigma }_{\rm{i}}})}}{{{\sigma _{\rm{i}}}^2}} = {e^{{y_i}^2}} - 1,\]
we see that our assumptions imply that
\[A = \left[ {\begin{array}{*{20}{c}}
   {1.28}  \\
   {0.73}  \\
\end{array}} \right]{\text{ and }}B = \left[ {\begin{array}{*{20}{c}}
   {1.03} & {1.11}  \\
   {1.11} & {1.31}  \\
\end{array}} \right]\]
We now run the following experiment:\\
Step 1: Initialize three accounts with \$1 of capital, i.e. ${W_{1,2,3}}(0)=\$1.$\\
Step 2: Draw ${\hat \mu _1},\;{\hat \mu _2},\;{\hat \sigma _1},\;{\hat \sigma _2}$ from the above distributions.\\
Step 3: Compute 
\[\begin{array}{l}
 {f_i}^{naive} = \frac{1}{{1 - x}}[{{\hat \Phi }^{ - 1}}\vec{\hat{\Delta}} ){]_i}\frac{{{{\hat S}_i}}}{{{{\hat \sigma }_i}}},\; \\ 
 {f_i}^{better} = \frac{1}{{1 - x}}{[{(\hat \Phi  * B)^{ - 1}}(\vec{\hat{\Delta}}  * \vec A)]_i}\frac{{{{\hat S}_i}}}{{{{\hat \sigma }_i}}},{\rm{ and }} \\ 
 {f_i}^{true} = \frac{1}{{1 - x}}[{\Phi ^{ - 1}}\vec \Delta ){]_i}\frac{{{S_i}}}{{{\sigma _i}}}. \\ 
 \end{array}\]
Step 4: Generate one-step returns $r_{1,2}$ for securities 1 and 2 using their true parameters and apply them to the holdings:
\[{{\rm{W}}_{\rm{1}}} \to {{\rm{W}}_{\rm{1}}}\exp (\sum\limits_i {{f_i}^{naive}{r_i}} ),\]
\[{{\rm{W}}_{\rm{2}}} \to {{\rm{W}}_{\rm{2}}}\exp (\sum\limits_i {{f_i}^{better}{r_i}} ),\]
\[{{\rm{W}}_{\rm{3}}} \to {{\rm{W}}_{\rm{3}}}\exp (\sum\limits_i {{f_i}^{true}{r_i}} ).\]
Step 5: Goto Step 2.
\\\\\noindent
We repeated this loop 100,000 times, then computed the three portfolio return series  ${{\rm{R}}_{\rm{j}}}(t) = \ln (\frac{{{W_j}(t)}}{{{W_j}(t - 1)}})$ and found:\\\\
Sharpe ratio$\left\{ {\;{{\rm{R}}_{\rm{1}}}(t)\;} \right\}$ = 0.27\\
Sharpe ratio$\left\{ {\;{{\rm{R}}_{\rm{2}}}(t)\;} \right\}$ = 0.37\\
Sharpe ratio$\left\{ {\;{{\rm{R}}_{\rm{3}}}(t)\;} \right\}$ = 0.46\\
\\
These are qualitatively as we expect:
\begin{enumerate}
\item The true parameters gave the expected result of about 0.47, up to sampling error.
\item The naive application of the standard mean-variance framework using the estimated parameters gives 0.27, a much worse result than the theoretically optimal 0.47 due to the negative convexity of the expected change in utility with respect to the investment fractions.
\item  The approach outlined in this paper gives 0.37, a result significantly better than 0.27 but still worse than 0.46.
\end{enumerate}
Finally, note that we've also run experiments in which we allowed the investor's estimates of her uncertainty to themselves be wrong, i.e. we've run the above experiment with ${f_i}^{better}$ computed using noisy $\vec{A}\text{ and }B$ (stemming from imperfect knowledge of how noisy are the parameter estimates).  The improvement in risk-adjusted performance is found to be very robust, i.e. substantial errors in $\vec{A}\text{ and }B$ do not appreciably diminish the performance and Sharpe ratio$\left\{ {\;{{\rm{R}}_{\rm{2}}}(t)\;} \right\}$ was well above Sharpe ratio$\left\{ {\;{{\rm{R}}_{\rm{1}}}(t)\;} \right\}$ in every experiment.
\theendnotes
\end{document}